\documentclass[11pt,a4paper]{article}
\usepackage{jcappub}
\usepackage{euscript,epsfig,amsmath,amssymb}
\usepackage{amsfonts,latexsym}

\newcommand{\be}[1]{\begin{equation}\label{#1}}
\newcommand{\ee}{\end{equation}}
\newcommand{\ba}[1]{\begin{eqnarray}\label{#1}}
\newcommand{\ea}{\end{eqnarray}}
\newcommand{\rf}[1]{(\ref{#1})}
\newcommand{\nn}{\nonumber}

\begin{document}

\title{Scalar perturbations in cosmological models with dark energy -- dark matter interaction}

\author{Maxim Eingorn$^{1,2}$,}
\author{Claus Kiefer$^{1}$}

\affiliation{$^{1}$Institute for Theoretical Physics, University of Cologne,\\ Z\"ulpicher st. 77, Cologne 50937, Germany\\}

\affiliation{$^{2}$Physics Department, North Carolina Central University,\\ Fayetteville st. 1801, Durham, North Carolina 27707, U.S.A.\\}

\emailAdd{maxim.eingorn@gmail.com} \emailAdd{kiefer@thp.uni-koeln.de}

\abstract{Scalar cosmological perturbations are investigated in the framework of a model with interacting dark energy and dark matter. In addition to
these constituents, the inhomogeneous Universe is supposed to be filled with the standard noninteracting constituents corresponding to the conventional
$\Lambda$CDM model. The interaction term is chosen in the form of a linear combination of dark sector energy densities with evolving coefficients. The
methods of discrete cosmology are applied, and strong theoretical constraints on the parameters of the model are derived. A brief comparison with
observational data is performed.}

\maketitle

\flushbottom

\section{Introduction}

The coincidence problem is among the greatest challenges in modern cosmology. It consists in the observational evidence that within the conventional
$\Lambda$CDM model the energy fractions of dark energy and dark matter are comparable (or, in other words, represent quantities of the same order) in the
Universe today, despite the fact that these constituents evolve in a quite different way; namely, the $\Lambda$-term does not evolve at all, while the energy
density of cold dark matter (CDM) decreases during the cosmological expansion. Dark energy represented by the $\Lambda$-term completely dominates the evolution
of the future Universe. However, at the early evolution stage it is the CDM that plays the leading part. In this connection the following natural question
arises: why are the dark sector contributions comparable today? Rejecting randomness as an explanation for such a coincidence, one can alleviate this challenge
by introducing direct non-gravitational dark energy -- dark matter interaction \cite{LiZhang,XuWang,SunYue,YangXu,Praseetha} (see also the important early
papers \cite{Amendola1,Amendola2} on coupled scalar fields and the detailed review \cite{CopelandSami} on dark energy dynamical behavior).
From the particle physics point of view, the case of uncoupled dark energy and dark matter is anyway a particular one, and in general their coupling may be
nonzero and natural \cite{YangXu}. As an example of such a coupling, let us mention the mass of CDM particles depending on the dark energy field
\cite{Bolotin}. It should be mentioned in addition that the introduction of dark sector interactions is also reasonable for the holographic scenario
\cite{Chimento,ZhangZhao}. Besides, the data of observations from the Abell Cluster A586 \cite{Arevalo,Pasqua,Bertolami}, measurements of the baryon acoustic
oscillations of the Lyman alpha forest from high redshift quasars \cite{AbdallaFerreira} as well as the cosmic microwave background and redshift-space
distortions measurements \cite{SalvatelliSaid} may indicate the reality of these interactions.

In the present paper, we use the methods of discrete cosmology devised in \cite{EZcosm1,EKZ2,EZcosm2} for the $\Lambda$CDM scenario to investigate a
cosmological model with dark energy -- dark matter interaction. Our approach consists in describing scalar perturbations at the stages of the cosmic evolution
when inhomogeneities such as galaxies and galaxy clusters have already been formed by representing the nonrelativistic pressureless matter (dust) as a system
of separate gravitating point-like particles. This is in accordance with the observational evidence that the typical distances between separate cosmic bodies
essentially exceed their dimensions. This well-grounded idea turns out to be interrelated with gravitational potentials and Newtonian equations of motion on
the FLRW background, concepts which are commonly used in modern $N$-body simulations. At the same time, it leads to theoretical as well as experimental
restrictions for numerous extensions of the conventional model and its alternatives. For example, methods of discrete cosmology have already been applied
successfully to cosmological models including a perfect fluid with a constant negative parameter in the linear equation of state (e.g., quintessence, the
phantom field, topological defects such as cosmic strings and domain walls) \cite{BEZ1}, Chevallier-Polarski-Linder (CPL) and other linear parametrizations of
the equation of state parameter \cite{Mariam}, quark gluon plasma that has escaped hadronization and survived up to now in the form of quark-gluon ``nuggets''
\cite{Jenk1,Jenk2}, nonlinear gravity \cite{Novak}, and the Lattice Universe topology \cite{Lattice}.

Here, we use the same approach with respect to the inhomogeneous Universe containing the standard $\Lambda$CDM constituents as well as interacting dark matter
and dark energy. We start by analyzing the equations of the scalar perturbations theory for the nonzero interaction term, then derive the conditions which must
be satisfied by its parameters, and conclude by briefly comparing the established theoretical limitations with the observations.

\section{Dark sector interactions and scalar cosmological perturbations}

First of all, let us write down the standard Friedmann equations describing the evolution of the homogeneous background in the framework of the
conventional $\Lambda$CDM model supplemented with the dark sector interactions:
\be{1} \frac{3\left(\mathcal{H}^2+\mathcal{K}\right)}{a^2}=\kappa \bar
T_0^0+\Lambda+\kappa\tilde\varepsilon_{DM}+\kappa\tilde\varepsilon_{DE}+\kappa\varepsilon_{rad}\, ,\ee
\be{2} \frac{2\mathcal{H}'+\mathcal{H}^2+\mathcal{K}}{a^2}=\Lambda-\kappa\tilde p_{DE}-\kappa p_{rad}\, .\ee
Here, $\mathcal{H}=a'/a$; the prime denotes the derivative with respect to the conformal time $\eta$; $\mathcal K=0,\pm1$ represents the spatial curvature;
$a(\eta)$ is the scale factor; $\kappa\equiv8\pi G_N/c^4$, where $G_N$ is the Newtonian gravitational constant, and $c$ is the speed of light; $\bar
T_0^0=\bar\rho c^2/a^3$ is the average energy density of the noninteracting nonrelativistic matter (both dark and visible), where $\bar\rho$ is the
corresponding average comoving rest mass density; $\Lambda$ represents the cosmological constant, while $\tilde\varepsilon_{DM}(\eta)$ and
$\tilde\varepsilon_{DE}(\eta)$ are the average energy densities of the (generally speaking) spacetime-dependent dark matter and dark energy, respectively,
interacting with each other; $\tilde p_{DE}$ is the pressure of the interacting dark energy (the pressure $\tilde p_{DM}$ of the interacting dark matter is
assumed to vanish, since it is also assumed to be completely nonrelativistic, similar to the noninteracting one); finally, $\varepsilon_{rad}(\eta)$ is the
average energy density of radiation, and $p_{rad}=\varepsilon_{rad}/3$ is the corresponding radiation pressure.

For simplicity and illustration, we focus our attention here on the vacuum-like equation of state $\tilde p_{DE}=-\tilde \varepsilon_{DE}$ for the interacting
dark energy, similar to the noninteracting one represented by the cosmological constant $\Lambda$ (see also, e.g., \cite{vacuum1,vacuum2}, where the same
choice is made). In this case, one can also use the notation $\kappa\tilde\varepsilon_{DE}=\tilde\Lambda$, and we can write $\kappa\tilde
p_{DE}=-\tilde\Lambda$. Here, evidently, $\tilde\Lambda(\eta)$ is not a constant; it varies with time in view of the presumptive non-gravitational interaction.

In the framework of the discrete cosmology (mechanical) approach to cosmological problems inside the cell of uniformity, developed recently in
\cite{EZcosm1,EKZ2,EZcosm2} (see also the recent papers \cite{Eleonora,Clarkson,Chisari,Ruth,EllGibb} on related issues), let us consider the following FLRW
metric perturbed by inhomogeneities:
\be{3} ds^2\approx a^2\left[(1+2\Phi)d\eta^2-(1-2\Phi)\gamma_{\alpha\beta}dx^{\alpha}dx^{\beta}\right],\quad \alpha,\beta=1,2,3\, .\ee

Here, $\Phi(\eta,{\bf r})=\varphi({\bf r})/\left(c^2a\right)$, where the introduced function $\varphi$ depends only on comoving spatial coordinates ${\bf r}$
and does not depend on time within the adopted accuracy (both the nonrelativistic and weak field limits are applied), in agreement with
\cite{Landau,Peebles,gadget2}. Really, this is a general solution of the equation $\Phi'+\mathcal{H}\Phi=0$, where contributions from peculiar velocities of
inhomogeneities are neglected. A noteworthy feature of this solution lies in the fact that for a single delta-shaped inhomogeneity it has the correct Newtonian
limit $\Phi\sim1/R$ when $\varphi\sim1/r$, where $R=ar$ is the ``physical'' distance (see also \cite{EZcosm1,EZcosm2} for details). The scalar perturbation
$\Phi$ satisfies the following system of linearized Einstein equations \cite{Mukhanov,Rubakov}:
\be{4} \triangle \Phi+3\mathcal{K} \Phi=\frac{1}{2}\kappa a^2\left(\delta
T_0^0+\delta\tilde\varepsilon_{DM}+\delta\tilde\varepsilon_{DE}+\delta\varepsilon_{rad}\right)\, ,\ee
\be{5} \Phi''+3\mathcal{H}\Phi' +\left(2\mathcal{H}'+\mathcal{H}^2\right)\Phi-\mathcal{K}\Phi=\frac{1}{2}\kappa a^2\left(\delta\tilde p_{DE}+\delta
p_{rad}\right)= \frac{1}{2}\kappa a^2\left(-\delta\tilde\varepsilon_{DE}+\frac{1}{3}\delta\varepsilon_{rad}\right)\, ,\ee
where $\triangle$ stands for the Laplace operator defined with respect to the spatial metric coefficients $\gamma_{\alpha\beta}$, and the fluctuation of the
energy density $T_0^0$ of the noninteracting nonrelativistic matter reads
\be{6} \delta T_0^0=\frac{\delta\rho({\bf r}) c^2}{a^3}+\frac{3\bar\rho c^2\Phi}{a^3}=\frac{\delta\rho({\bf r}) c^2}{a^3}+\frac{3\bar\rho\varphi({\bf
r})}{a^4}\, .\ee
Here, $\delta\rho({\bf r})$ stands for the fluctuation of the rest mass density in comoving coordinates. From \rf{5} with the help of \rf{1} and \rf{2} we
immediately obtain
\be{8} \frac{\bar\rho\varphi({\bf r})}{a^4}+\tilde\varepsilon_{DM}(\eta)\frac{\varphi({\bf r})}{c^2a}+\frac{4}{3}\varepsilon_{rad}(\eta)\frac{\varphi({\bf
r})}{c^2a}= \delta\tilde\varepsilon_{DE}-\frac{1}{3}\delta\varepsilon_{rad}\, .\ee

It is worth noting that this equation agrees with the corresponding equation (40) in \cite{Vlasov} in the absence of interacting dark constituents. The
third term on the left-hand side (lhs) of \rf{8} behaves as $1/a^5$, which is outside the limits of our adopted accuracy and therefore should be dropped,
while the first term on the lhs as well as the second term on the right hand side (rhs) both behave as $1/a^4$. In other words, the background radiation
contribution can be disregarded. Then we get from \rf{8}:
\be{9} \delta\tilde\varepsilon_{DE}=\frac{\bar\rho\varphi({\bf r})}{a^4}+\tilde\varepsilon_{DM}(\eta)\frac{\varphi({\bf
r})}{c^2a}+\frac{1}{3}\delta\varepsilon_{rad}\, .\ee

The dependence of the second term on the rhs of \rf{9} on the scale factor $a$ is different from $1/a^4$, since the quantity $\tilde\varepsilon_{DM}(\eta)$
does not behave as $1/a^3$ (this behavior is inconsistent with the dark sector interactions). Moreover, this term cannot be neglected because the cosmological
model under consideration includes interacting dark components with the energy densities comparable with the energy densities of noninteracting constituents at
present, as well as in the (at least) near past and in the (at least) near future. Consequently, even if $\delta\varepsilon_{rad}=-3\bar\rho\varphi({\bf
r})/a^4$, and the first and third terms on the rhs thus exactly compensate each other, we would still have $\delta\tilde\varepsilon_{DE}\neq0$; the interacting
dark energy thus cannot be homogeneous, in contrast to the noninteracting one represented by $\Lambda=\mathrm{const}$.

The substitution of \rf{9} into \rf{4} leads to the following equation:
\be{10} \triangle\varphi({\bf r})+3\mathcal{K}\varphi({\bf r})=\frac{1}{2}\kappa c^2\left(\delta\rho({\bf r}) c^2+\frac{4\bar\rho\varphi({\bf
r})}{a}+a^3\delta\tilde\varepsilon_{DM}+ a^2\tilde\varepsilon_{DM}(\eta)\frac{\varphi({\bf r})}{c^2}+\frac{4}{3}a^3\delta\varepsilon_{rad}\right)\, ,\ee
whence
\be{11} \triangle\varphi({\bf r})+3\mathcal{K}\varphi({\bf r})=\frac{1}{2}\kappa c^2\left[\delta\rho({\bf r}) c^2+f({\bf r})\right],\quad
\delta\tilde\varepsilon_{DM}=-\frac{4\bar\rho\varphi({\bf r})}{a^4}-\tilde\varepsilon_{DM}(\eta)\frac{\varphi({\bf
r})}{c^2a}-\frac{4}{3}\delta\varepsilon_{rad}+\frac{f({\bf r})}{a^3}\, .\ee
Here, the introduced function $f$ depends only on comoving spatial coordinates ${\bf r}$ similar to the function $\varphi$, while its dependence on time
should be disregarded.

It is worth mentioning that, as one can easily verify, the derived fluctuations \rf{9} and \rf{11} satisfy the energy conservation equation
\be{12} \delta\tilde\varepsilon_{tot}'+3\mathcal{H}(\delta\tilde\varepsilon_{tot}+\delta\tilde p_{tot})+(\tilde\varepsilon_{tot}+\tilde p_{tot})(-3\Phi')=0\,
,\ee
where the total energy density and pressure of the interacting dark components read, respectively,
$\tilde\varepsilon_{tot}(\eta)=\tilde\varepsilon_{DE}+\tilde\varepsilon_{DM}$ and $\tilde p_{tot}(\eta)=\tilde p_{DE}+\tilde p_{DM}=-\tilde\varepsilon_{DE}$.

The supposed interaction between the dark constituents of the Universe is usually described by the following equations
\cite{LiZhang,ZhangLiu,Costa,Bolotin,Praseetha,Arevalo,XuWang,SunYue,Pasqua,CaoLiang,YangXu,Li2Zhang}:
\be{14} \tilde\varepsilon_{DE}'+3\mathcal{H}(\tilde\varepsilon_{DE}+\tilde p_{DE})=\tilde\varepsilon_{DE}'=-Q\, ,\ee
\be{15} \tilde\varepsilon_{DM}'+3\mathcal{H}(\tilde\varepsilon_{DM}+\tilde p_{DM})=\tilde\varepsilon_{DM}'+3\mathcal{H}\tilde\varepsilon_{DM}=Q\, ,\ee
where the phenomenological interaction term $Q$ often represents a linear combination of $\tilde\varepsilon_{DE}$ and $\tilde\varepsilon_{DM}$ with some
(generally speaking) time-varying coefficients $\Gamma_1$ and $\Gamma_2$, which depend neither on $\tilde\varepsilon_{DE}$, nor on
$\tilde\varepsilon_{DM}$, and may change in time only because of some postulated reaction rate evolution of unknown nature:
\be{16} Q=\Gamma_1\tilde\varepsilon_{DE}+\Gamma_2\tilde\varepsilon_{DM}\, .\ee
Consequently, the fluctuations must satisfy the appropriate perturbed equations:
\be{17} \delta\tilde\varepsilon_{DE}'+3\mathcal{H}(\delta\tilde\varepsilon_{DE}+\delta\tilde p_{DE})+(\tilde\varepsilon_{DE}+\tilde
p_{DE})(-3\Phi')=\delta\tilde\varepsilon_{DE}'=-\delta Q\, ,\ee
\be{18} \delta\tilde\varepsilon_{DM}'+3\mathcal{H}(\delta\tilde\varepsilon_{DM}+\delta\tilde p_{DM})+(\tilde\varepsilon_{DM}+\tilde
p_{DM})(-3\Phi')=\delta\tilde\varepsilon_{DM}'+3\mathcal{H}\delta\tilde\varepsilon_{DM}+\tilde\varepsilon_{DM}(-3\Phi')=\delta Q\, ,\ee
where
\be{19} \delta Q=\Gamma_1\delta\tilde\varepsilon_{DE}+\Gamma_2\delta\tilde\varepsilon_{DM}\, .\ee

Hereinafter, we focus our attention on the most popular case of the following parametrization
\cite{LiZhang,ZhangLiu,Costa,Bolotin,Praseetha,XuWang,SunYue,Pasqua,CaoLiang,YangXu,Li2Zhang}:
\be{23} \Gamma_1=\mathcal{H}\gamma_1,\quad \Gamma_2=\mathcal{H}\gamma_2\, ,\ee
where $\gamma_1$ and $\gamma_2$ are constant (time-independent) parameters. It is important to stress that the quantity $\mathcal{H}$ is treated here as
characterizing the global average expansion rate. Consequently, neither $\mathcal{H}$ nor $\Gamma_{1,2}$ are perturbed when passing from \rf{16} to \rf{19}. At
the same time, according to the reasoning presented in \cite{deltaH1} (see also \cite{deltaH2,deltaH3}), one should not ignore the perturbation $\delta H$
describing the difference between the local expansion rate and the global one, which is defined by means of the relationship
\be{h1} u^i_{;i}\approx\frac{3}{a}(\mathcal{H}+\delta\mathcal{H})\, ,\ee
where $u^i$, $i=0,1,2,3$, stand for the $4$-velocity components. However, it is easy to demonstrate that within the adopted accuracy $\delta\mathcal{H}=0$.
Really, since we neglect the peculiar motion of inhomogeneities, the only nonzero component of the $4$-velocity is $u^0\approx(1-\Phi)/a$. Then the calculation of
the covariant derivative on the lhs of Eq.~\rf{h1} on the basis of the metric \rf{3} gives
\be{h2}
u^i_{;i}\approx\left(\frac{1-\Phi}{a}\right)'+\frac{4a'}{a}\frac{1-\Phi}{a}-\frac{2\Phi'}{a}=\frac{3\mathcal{H}}{a}-\frac{3\mathcal{H}\Phi}{a}-\frac{3\Phi'}{a}\ee
up to the first order of smallness. Further, taking into account that $\Phi(\eta,{\bf r})=\varphi({\bf r})/\left(c^2a\right)$, we conclude that
$u^i_{;i}\approx3\mathcal{H}/a$. Then the comparison with \rf{h1} immediately gives $\delta\mathcal{H}=0$. Thus, disregard of the perturbation $\delta\mathcal{H}$
is really justified within the accuracy of our approach.

Substituting \rf{19} and \rf{23} into \rf{17}, we obtain
\be{20} \delta\tilde\varepsilon_{DE}'=-\mathcal{H}\left(\gamma_1\delta\tilde\varepsilon_{DE}+\gamma_2\delta\tilde\varepsilon_{DM}\right)\, .\ee
Furthermore, substituting \rf{9} and \rf{11} into \rf{20}, we get after some computation
\be{25} \gamma_1\tilde\varepsilon_{DE}(\eta)=(4-\gamma_1+4\gamma_2)\frac{\bar\rho c^2}{a^3}+(4-\gamma_1)\tilde\varepsilon_{DM}(\eta)+ \frac{c^2a}{3\varphi({\bf
r})}(4-\gamma_1+4\gamma_2)\delta\varepsilon_{rad}-\gamma_2\frac{c^2f({\bf r})}{a^2 \varphi({\bf r})}\, .\ee

In view of \rf{12}, Eq. \rf{18} gives nothing new and may be dropped. The substitution of \rf{16} and \rf{23} into \rf{14} and \rf{15} leads to the
equations
\be{2627} \frac{d\tilde\varepsilon_{DE}}{da}=-\frac{1}{a}(\gamma_1\tilde\varepsilon_{DE}+\gamma_2\tilde\varepsilon_{DM}),\quad
\frac{d\tilde\varepsilon_{DM}}{da}=\frac{1}{a}[\gamma_1\tilde\varepsilon_{DE}+(\gamma_2-3)\tilde\varepsilon_{DM}]\, .\ee

After some computation with the help of \rf{25} and \rf{2627}, one can show that
\be{32} -(4-\gamma_1+4\gamma_2)\tilde\varepsilon_{DM}(\eta)=(4-\gamma_1+4\gamma_2)\frac{\bar\rho c^2}{a^3}+ \frac{c^2a}{3\varphi({\bf
r})}(4-\gamma_1+4\gamma_2)\delta\varepsilon_{rad}-2\gamma_2\frac{c^2f({\bf r})}{a^2\varphi({\bf r})}\, .\ee

First, let us consider the case $4-\gamma_1+4\gamma_2=0$, which gives $\gamma_2f({\bf r})=0$. If $\gamma_2=0$, and thus $\gamma_1=4$, there is a contradiction
with Eq. \rf{25}. Therefore, $\gamma_2\neq0$, while $f({\bf r})=0$. According to \rf{11}, in this case the nonrelativistic gravitational potential is
determined only by inhomogeneities of noninteracting matter as in \cite{EZcosm1,EKZ2,EZcosm2}. From \rf{25} and \rf{2627} we obtain
\be{a} \tilde\varepsilon_{DE}(\eta)=Aa^{-3(1+\gamma_2)},\quad \tilde\varepsilon_{DM}(\eta)=-\left(1+\frac{1}{\gamma_2}\right)Aa^{-3(1+\gamma_2)},\quad
A=\mathrm{const}\, .\ee

If we demand in addition that both energy densities in \rf{a} are positive, then $A>0$ and $-1<\gamma_2<0$, so $\tilde\varepsilon_{DE}(\eta)$ and
$\tilde\varepsilon_{DM}(\eta)$ decrease in the expanding Universe in view of the inequality $3(1+\gamma_2)>0$. According to \rf{9} and \rf{11}, the same
statement holds true for their fluctuations. Besides, $Q>0$, and this means an energy transfer from dark energy to dark matter. Exactly this transfer direction
is required by the second law of thermodynamics as well as for solving the coincidence problem \cite{Costa,XuWang,Pasqua,Mazumder,WangLin,PavonWang}. It is
worth noting that if the parameters $\gamma_1$ and $\gamma_2$ are supposed to be equal, then we have $\gamma_1=\gamma_2=-4/3$, so the condition $\gamma_2>-1$
does not hold true.

In the second case we have $4-\gamma_1+4\gamma_2\neq0$. Therefore, from \rf{32} it immediately follows that
\be{33} \delta\varepsilon_{rad}=-\frac{3\bar\rho\varphi({\bf r})}{a^4},\quad \tilde\varepsilon_{DM}(\eta)=\frac{2\gamma_2}{4-\gamma_1+4\gamma_2}\frac{c^2f({\bf
r})}{a^2\varphi({\bf r})}\, .\ee

Here, we take into account the above-mentioned fact that the quantity $\tilde\varepsilon_{DM}(\eta)$ does not behave as $1/a^3$, so the only possibility to
compensate the last term on the rhs of Eq. \rf{32} is to require that $\tilde\varepsilon_{DM}(\eta)\sim1/a^2$ and $\gamma_2f({\bf r})\neq0$. From \rf{25},
\rf{2627} and \rf{33} we get the equality $2-\gamma_1-2\gamma_2=0$ and the expressions
\be{b} \tilde\varepsilon_{DM}(\eta)=\frac{\gamma_2}{1+3\gamma_2}\frac{c^2f({\bf r})}{a^2\varphi({\bf r})},\quad
\tilde\varepsilon_{DE}(\eta)=\frac{1}{2}\tilde\varepsilon_{DM}=\frac{\gamma_2}{2(1+3\gamma_2)}\frac{c^2f({\bf r})}{a^2\varphi({\bf r})}\, .\ee

As in the previous case, both energy densities $\tilde\varepsilon_{DE}(\eta)$ and $\tilde\varepsilon_{DM}(\eta)$ decrease in the expanding Universe together
with their fluctuations. Besides, we again demand additionally that $\tilde\varepsilon_{DE,DM}>0$ and come to the conclusion that the energy transfer from dark
energy to dark matter takes place, as before, in view of the inequality $Q>0$. It is also worth mentioning that if the parameters $\gamma_1$ and $\gamma_2$ are
equal to each other, then their common value is $2/3$.

Expressing the function $f({\bf r})$ from \rf{b} and substituting the result into \rf{11}, we derive the following equation for the nonrelativistic
gravitational potential:
\be{c} \triangle\varphi({\bf r})+3\mathcal{K}\varphi({\bf r})-\lambda\varphi({\bf r})=\frac{1}{2}\kappa c^4\delta\rho({\bf r})=4\pi G_N[\rho({\bf
r})-\bar\rho],\quad \lambda=\frac{1+3\gamma_2}{2\gamma_2}\kappa a^2\tilde\varepsilon_{DM}(\eta)=\mathrm{const}\, . \ee

For simplicity, let us restrict ourselves to the case of flat spatial geometry ($\mathcal{K}=0$) and require that $\lambda>0$, so either $\gamma_2<-1/3$ or
$\gamma_2>0$. Then the solution of Eq. \rf{c} for a system of gravitating masses $m_i$ with the comoving radius-vectors ${\bf r}_i$, characterized by the rest
mass density
\be{d} \rho({\bf r})=\sum\limits_i m_i\delta({\bf r}-{\bf r}_i)\, ,\ee
reads
\be{e} \varphi({\bf r})=-G_N\sum\limits_i \frac{m_i}{|{\bf r}-{\bf r}_i|}\exp\left(-\sqrt{\lambda}|{\bf r}-{\bf r}_i|\right)+\frac{4\pi G_N\bar\rho}{\lambda}\,
.\ee

It is very important that this function is characterized by the zero average value $\bar\varphi=0$ for any mass distribution, in contrast to the potential
determined within the bounds of the conventional $\Lambda$CDM model, which may have a nonzero value after performing a procedure of spatial averaging (for
an example of the corresponding mass distribution, see \cite{EZcosm2,EZcosm1,EBV}).

\section{Comparison with experimental restrictions, and generalization}

Summarizing, we have arrived at the following two allowed cases:

\

{\bf I. \ $4-\gamma_1+4\gamma_2=0$, \ $f=0$, \ $-1<\gamma_2<0$, \ $A>0$}

{\bf II. \ $2-\gamma_1-2\gamma_2=0$, \ $f\gamma_2(1+3\gamma_2)>0$}

\

Provided that the most widespread case of flat spatial geometry ($\mathcal{K}=0$) is studied, the second class becomes even more constrained:

\

{\bf II. \ $2-\gamma_1-2\gamma_2=0$, \ $f>0$, \ $\gamma_2<-1/3$ or $\gamma_2>0$}

\

Now let us illustrate these constraints with three popular examples:

\

{\bf A. \ $Q=\mathcal{H}\gamma_1\tilde\varepsilon_{DE}$ ($\gamma_2=0$)

B. \ $Q=\mathcal{H}\gamma_2\tilde\varepsilon_{DM}$ ($\gamma_1=0$)

C. \ $Q=\mathcal{H}\gamma\left(\tilde\varepsilon_{DE}+\tilde\varepsilon_{DM}\right)$ ($\gamma_1=\gamma_2\equiv\gamma$)}

\

One can easily verify that the model {\bf A} satisfies neither the first set of conditions ({\bf I}), nor the second one ({\bf II}). The models {\bf B} and
{\bf C} also do not satisfy the conditions {\bf I}, while the conditions {\bf II} lead to the following allowed interaction terms, respectively:
$Q=\mathcal{H}\tilde\varepsilon_{DM}$ and $Q=(2/3)\mathcal{H}\left(\tilde\varepsilon_{DE}+\tilde\varepsilon_{DM}\right)$. If the noninteracting dark sector
components were absent, both expressions would be incompatible with known observational restrictions for the parameters $\gamma_1$ and $\gamma_2$, see
\cite{ZhangLiu,Costa,Bolotin,Pasqua,CaoLiang}. Consequently, two of the above-mentioned interaction terms are consistent with the theory of scalar cosmological
perturbations and with observational data only if in addition to interacting dark constituents of the inhomogeneous Universe there are noninteracting
constituents weakening the empirical limitations. Actually, this is the reason why we have included the noninteracting dark sector components (in particular,
the constant $\Lambda$-term) in the model. In other words, after imposing theoretical constraints, the interacting dark energy does not necessarily ensure,
e.g., the late-time acceleration of the Universe expansion (in this case the ``dark energy'' notion itself loses the initial sense). Therefore, the constant
$\Lambda$-term may be still needed for reaching agreement with the supernovae and other data. Thus, it is reasonable not to exclude such possible contributions
from the very beginning.

The finding that three popular interaction terms quoted above are in general incompatible with the derived constraints actually originates from Eq.~\rf{25} and
is certainly valid for the investigated simplest case of the vacuum-like equation of state $\tilde p_{DE}=-\tilde \varepsilon_{DE}$. The more general and
complicated case of an arbitrary constant parameter $\omega$ in the linear equation of state $\tilde p_{DE}=\omega\tilde \varepsilon_{DE}$ lies beyond the
scope of our current investigation, representing a subject of a separate forthcoming paper. However, let us laconically generalize Eq.~\rf{25} for $\omega<-1$
or $-1<\omega<0$. First, instead of \rf{9} and \rf{11} now we have, respectively,
\be{add1} \delta\tilde\varepsilon_{DE}=-\frac{\bar\rho\varphi({\bf r})}{\omega a^4}-\tilde\varepsilon_{DM}(\eta)\frac{\varphi({\bf r})}{\omega
c^2a}-\frac{1}{3\omega}\delta\varepsilon_{rad}-(1+\omega)\tilde\varepsilon_{DE}(\eta)\frac{\varphi({\bf r})}{\omega c^2a}\, ,\ee
\be{add2} \delta\tilde\varepsilon_{DM}=-\left(3-\frac{1}{\omega}\right)\frac{\bar\rho\varphi({\bf
r})}{a^4}+\frac{1}{\omega}\tilde\varepsilon_{DM}(\eta)\frac{\varphi({\bf
r})}{c^2a}-\frac{1}{3}\left(3-\frac{1}{\omega}\right)\delta\varepsilon_{rad}+\frac{f({\bf r})}{a^3}+(1+\omega)\tilde\varepsilon_{DE}(\eta)\frac{\varphi({\bf
r})}{\omega c^2a}\, .\ee

Second, the background energy densities $\tilde\varepsilon_{DE}(\eta)$ and $\tilde\varepsilon_{DM}(\eta)$ now satisfy the equations
\be{add3} \frac{d\tilde\varepsilon_{DE}}{da}=-\frac{1}{a}[(\gamma_1+3+3\omega)\tilde\varepsilon_{DE}+\gamma_2\tilde\varepsilon_{DM}],\quad
\frac{d\tilde\varepsilon_{DM}}{da}=\frac{1}{a}[\gamma_1\tilde\varepsilon_{DE}+(\gamma_2-3)\tilde\varepsilon_{DM}]\, .\ee

Finally, the generalization of Eq.~\rf{25} reads:
\ba{add4} &{}&[\gamma_1-(1+\omega)\gamma_2-(1+\omega)(1+3\omega)]\tilde\varepsilon_{DE}(\eta)\nn\\
&=& [1-3\omega-\gamma_1+(1-3\omega)\gamma_2]\frac{\bar\rho c^2}{a^3}+[1-3\omega-\gamma_1+(1+\omega)\gamma_2)]\tilde\varepsilon_{DM}(\eta)\nn\\
&+& [1-3\omega-\gamma_1+(1-3\omega)\gamma_2]\frac{c^2a}{3\varphi({\bf r})}\delta\varepsilon_{rad}+\omega\gamma_2\frac{c^2f({\bf r})}{a^2 \varphi({\bf r})}\,
.\ea

The derived condition \rf{add4} should be taken into account along with Eqs.~\rf{add3} and plays the role of the additional restriction for the set of three
initially free parameters $\omega$, $\gamma_1$ and $\gamma_2$.

\section{Conclusion}

In the present paper, we have investigated the scalar perturbations in the framework of the cosmological model with dark energy -- dark matter interaction. The
inhomogeneous Universe is supposed to be filled with the standard noninteracting constituents of the conventional $\Lambda$CDM model as well as with
interacting dark matter and dark energy (with dust-like and vacuum-like equations of state, respectively). The interaction term is taken in the frequently used
form \rf{16} with the coefficients \rf{23}. We have applied methods of the discrete cosmology approach in order to impose theoretical constraints on the model
under consideration. These constraints originate from the derived condition \rf{25} and are generally incompatible with the experimental data. Then we have
generalized this condition to the case of an arbitrary parameter $\omega$ in the linear equation of state of interacting dark energy. As a result, it takes the
form \rf{add4} and can be used again for restricting the parameters entering the interaction term, as well as the equation of state parameter $\omega$ itself.

\section*{Acknowledgements}

We would like to thank the Referees for valuable comments and critical remarks which have considerably improved the presentation of the obtained results. We
appreciate useful discussions with A.~Zhuk as well. The work of M. Eingorn was supported partially by NSF CREST award HRD-1345219 and NASA grant NNX09AV07A. M.
Eingorn also thanks the German Academic Exchange Service (DAAD) for the scholarship during his research visit to Cologne.

\end{document}